\documentclass[prd,showpacs,preprintnumbers,floatfix,superscriptaddress,twocolumn]{revtex4}

\usepackage{amssymb}
\usepackage[pdftex]{color,graphicx}

\newcommand{\gev}{\,{\rm GeV}}

\newcommand{\fm}{\,{\rm fm}}
\newcommand{\mpi}{m_\pi}

\newcommand{\del}{\partial}

\newcommand{\chisq}{\chi^2_{\rm dof}}

\begin{document}

\preprint{ANL-PHY-12245-THY-2009, JLAB-THY-08-922}
\title{Octet baryon masses and sigma terms from an SU(3) chiral extrapolation}

\author{R.~D.~Young}
\affiliation{Physics Division, Argonne National Laboratory, Argonne, Illinois 60439, USA}
\author{A.~W.~Thomas}
\affiliation{Jefferson Lab, 12000 Jefferson Ave., Newport News, Virginia 23606, USA}
\affiliation{College of William and Mary, Williamsburg, Virginia 23187, USA}

\date{\today}

\begin{abstract}
  We report an analysis of the impressive new lattice simulation
  results for octet baryon masses in 2+1-flavor QCD. The analysis is
  based on a low order expansion about the chiral SU(3) limit in which
  the symmetry breaking arises from terms linear in the quark masses
  plus the variation of the Goldstone boson masses in the leading
  chiral loops. The baryon masses evaluated at the physical light
  quark masses are in remarkable agreement with the experimental
  values, with a model dependence considerably smaller than the rather
  small statistical uncertainty. From the mass formulae one can
  evaluate the sigma commutators for all octet baryons. This yields an
  accurate value for the pion-nucleon sigma commutator.  It also
  yields the first determination of the strangeness sigma term based
  on 2+1-flavor lattice QCD and, in general, the sigma commutators
  provide a resolution to the difficult issue of fine-tuning the
  strange quark mass.
\end{abstract}

%\pacs{}

\maketitle

%%%%%%%%%%%%%%%%%%%%%%%%%%%%%%%%%%%%%%%%%%%%%%%%%%%%%%%%%%%%%
%
In recent years lattice QCD has matured to a level where it can be
used as a precision tool to confront experimental aspects of
nonperturbative QCD. This advance has occured primarily in the
heavy-meson sector \cite{Davies:2003ik}, while for light-quarks,
and especially for light-quark baryons, progress has been more
steady. However, new high-precision studies of the baryon spectrum
in 2+1-flavour dynamical simulations have recently been reported by LHPC
\cite{WalkerLoud:2008bp}, PACS-CS \cite{Aoki:2008sm}, HSC
\cite{Lin:2008pr} and D\"urr {\em et al.} \cite{Durr:2008zz}. 
In this Letter, we demonstrate 
that these results are sufficiently close to the chiral regime that they permit
a reliable determination of the octet baryon masses at the physical point, as well 
the associated sigma terms.

Our analysis focusses on the (chronologically) first two of the above
cited studies, those of LHPC and PACS-CS. The smallest pion masses are
of order $0.29\gev$ and $0.30\gev$, respectively. The lattice spacing
for the MILC configurations, used by LHPC, has been carefully
determined in heavy-quark systems \cite{Aubin:2004wf}. That same
analysis also produced a physical determination of the Sommer scale,
$r_0=0.465\pm 0.012\fm$, which we use to set the scale in the PACS-CS
simulations.

Our working hypothesis is that the (different) improved actions
employed by the two groups yield a very good approximation to the
continuum theory, so that discretization artefacts are small. A
unified treatment of the continuum extrapolation is only possible once
there are results available at multiple lattice spacings, such as the
study of Ref.~\cite{Durr:2008zz}. On the other hand, a comparison of
the absolute values of the baryon masses extracted from the analysis
of the separate LHPC and PACS-CS data sets provides a test, {\it a
  posteriori}, of the validity of this hypothesis. We shall see that
it appears to be a very good approximation.

The first step in our analysis is to calculate the finite-volume
corrections within the low-energy effective field theory (EFT), at
each combination of quark masses. We can then focus our attention on
the extrapolation within an infinite-volume, continuum effective field
theory framework. A detailed volume dependence analysis of the nucleon
mass in 2-flavour simulations has observed that the finite-volume
corrections are well-described by the leading one-loop results of
chiral effective field theory \cite{Ali Khan:2003cu}. We extend this
approach to the SU(3) case in order to estimate the infinite-volume
limit of the present lattice results. While the leading infrared
effects of the loop integral are independent of the ultraviolet
regularisation, to be conservative we include an uncertainty in this
correction which amounts to the difference between the finite-volume
correction determined without a regulator \cite{Beane:2004tw} and that
evaluated with a dipole of mass of $0.8\gev$ \cite{Young:2002cj}.  The
largest shift we use in our analysis is for the nucleon at the
lightest pion mass (where $\mpi L=3.7$), giving a correction of
$-0.019\pm 0.005\gev$.

The chiral expansion of the octet baryons has been presented on
numerous occasions in the literature --- eg.,
Refs.~\cite{Borasoy:1996bx,Donoghue:1998bs,WalkerLoud:2004hf,Frink:2004ic}.
It may be expressed as
\begin{equation}
M_B = M^{(0)} + \delta M_B^{(1)} + \delta M_B^{(3/2)} + \ldots\,,
\label{eq:MB32}
\end{equation}
where the superscript denotes the order of the expansion in powers of
the quark mass --- the {\em explicit} chiral symmetry breaking
parameter of QCD. Here, $M^{(0)}$ denotes the baryon mass in the SU(3)
chiral limit.

The $\delta M_B^{(1)}$ term is linear in 
the scale-dependent, running quark masses, $m_l$ and $m_s$, we refer to
Ref.~\cite{WalkerLoud:2004hf} for explicit expressions.  
In this analysis, we replace the quark masses by the corresponding 
meson masses squared, as also done in
Ref.~\cite{WalkerLoud:2008bp}. To the order that we work in this
manuscript, the use of either quark masses or meson masses squared is
equivalent.
%One should caution the use of this substitution upon
%extending to higher order~\cite{McGovern:2006fm}.

The next higher order term, $\delta M_B^{(3/2)}$, involves the leading
one-loop corrections arising from the baryons coupling to the
pseudo-Goldstone bosons, $\phi\equiv \pi,\,K,\,\eta$.  For the
baryons, we include both the octet and decuplet. Certainly, in the
limit $m_\phi\ll \delta$ (the octet-decuplet mass-splitting in the
chiral limit), the decuplet contributions are formally of higher
order. However, with a physical kaon mass $\sim 0.7\gev$, it is clear
that $\delta\sim 0.3\gev$ cannot be treated as a heavy energy
scale. For this reason our formal assessment of the order of a given
diagram treats the octet and decuplet baryons as degenerate. However,
in order to more accurately represent the branch structure in the
transition region $m_\phi\sim\delta$, in the numerical evaluation of
the loop integrals we maintain the octet-decuplet mass
splitting. Further, the renormalisation is performed such that the
decuplet is formally integrated out in the limit $m_\phi\ll\delta$.

For the explicit forms of the loop integrals, we refer the reader to
Ref.~\cite{Young:2002ib}.  The coefficients of these loop
contributions \cite{WalkerLoud:2004hf} are expressed in terms of the
pseudoscalar decay constant and the relevant baryon axial charges. The
nucleon axial charge, $g_A=D+F=1.27$ is fixed by experiment, while all
other couplings are determined by SU(6) relations ($F=\frac23 D$ and
$\mathcal{C}=-2D$). We note that $\mathcal{C}$ is also related to the
decay width of the $\Delta$, from which a similar value can be inferred. We
also adopt the chiral perturbation theory estimate for the meson decay
constant in the SU(3) chiral limit, $f=0.0871\gev$
\cite{Amoros:2001cp}. The octet-decuplet splitting is chosen
phenomenologically to be the physical $N$--$\Delta$ splitting,
$\delta=0.292\gev$. In principle, all of these input parameters could
be constrained by actual lattice simulation results, at least in the
near future --- we refer to recent progress in the computation of the
%\cite{Edwards:2005ym,Yamazaki:2008py,Khan:2006de,Lin:2007ap,Alexandrou:2006mc}.
axial charges~\cite{axial}. For this study we take these
parameters from phenomenology, we incorporate generous uncertainties
in these inputs into our systematic error analysis, allowing $f$ to
each vary by $\pm5\%$, and $D$, $F$, $\cal C$ and $\delta$ to vary by
$\pm15\%$.  We leave a global analysis, that determines all these
inputs simultaneously from the same lattice calculation to future
work.

In fitting the lattice results, we wish to minimize the uncertainty
from higher-order terms in the chiral expansion. We therefore limit
ourselves to the smallest domain of the light-quarks possible with
these latest lattice results. For both the LHPC and PACS-CS results,
we include {\em only} the results of the simulations for
$\mpi^2<0.2\gev^2$.

We first fit the baryon masses to leading-order, keeping only $\delta
M_B^{(1)}$ in the expansion~(\ref{eq:MB32}). This gives a reasonable
description of the (finite-volume corrected) lattice results, with a
reduced $\chi^2$ ($\chisq$) of 0.9 and 0.3 for the LHPC and PACS-CS
results, respectively. We caution reading too much into the absolute
value of this naive $\chi^2$, as there are certainly strong
correlations among the points on the same gauge configurations, as
mentioned in Ref.~\cite{WalkerLoud:2008bp}. In particular, the
mass-splittings among the baryons will typically be known better than
one would estimate from naively adding the uncertainties in the
absolute masses.  Without further information on these correlations,
our $\chi^2$ wil typically underestimate of the results
constructed from the full information contained in the lattice
simulation.

We now investigate the inclusion of the loop corrections.
%Provided the meson masses are sufficiently
%light, the results will be independent of the ultraviolet treatment of
%the loop contributions. In this domain one maintains mathematical
%control and the model independence of the expansion is
%guaranteed. 
We perform fits where the regularisation scale-dependence has been
removed to {\em all} orders and just the leading term of the loop
integration is retained. Using the phenomenological coefficients, the
best fit produces a (naive) $\chisq$ of order the order 40 (36) for
the LHPC (PACS-CS) results.  Alternatively, a similar fit was also
done by the LHPC \cite{WalkerLoud:2008bp}, where a suitable $\chisq$
was acheived by allowing the chiral coefficients to be fit. While this
produced a reasonable description of the data, the determined
coefficients (such as $g_A^2$ or ${\cal C}^2$) were found to be an order of
magnitude smaller than phenomenological estimates.  The most natural
conclusion of these observations (as also noted in
\cite{WalkerLoud:2008bp}) is that the meson masses lie beyond the
model-independent or ``power counting'' region (PCR) of the expansion
at this order.
It is therefore a challenge to maintain the constraints of both the
EFT and the lattice results (without abandoning one of them).

The breakdown of the expansion at this order should come as no
surprise, as it has long been known that the SU(3) chiral expansion in
the baryon sector is quite poor --- even at the physical quark masses
(see eg.~\cite{Borasoy:1996bx,Donoghue:1998bs}). Donoghue, Holstein
and Borasoy formulated a solution to this problem through
long-distance regularisation (LDR)~\cite{Donoghue:1998bs}.
Concurrently, the same techniques were being developed to alleviate
the problem of chiral extrapolation at moderate quark masses in
lattice QCD with the development of finite-range regularisation
(FRR)~\cite{Leinweber:1999ig}. While this has largely been developed
for applications in SU(2) chiral
extrapolations~\cite{Leinweber:2003dg}, here we utilise the features
of the LDR/FRR formalism to perform extrapolations in the framework of
chiral SU(3).

For the relevant formulae describing the renormalisation of the loop
integrals, we direct the interested reader to
Ref.~\cite{Young:2002ib}.  Upon renormalisation, and to the order we
are working, the FRR forms produce exactly the same expansion as we
have described above.  The difference lies in a resummation of
higher-order terms, which are suppressed by inverse powers of the
regularisation scale. At this order, the lowest-order addition to the
renormalised expansion will explicitly appear in the form
$m_\phi^4/\Lambda$, with $\Lambda$ the regulator mass.

In working towards {\em ab initio} studies of QCD,
where no external, phenomenological input is used, we do not wish to
impose any external constraints on the regularisation
scale. Based on the success of FRR, we should like to
utilise the observation that the induced resummation of higher-order
terms provides an improved description of a range of
lattice obesrvables.  Here we use the lattice results themselves to
determine the regularisation scale by minimizing the $\chi^2$.

Some significant features of using the $\chi^2$ measure, with
the regularisation scale treated as a free parameter, are:
\begin{itemize}
\item no bias of a preferred scale is dictated by phenomenology;
\item if one is working with results that genuinely lie within the
  PCR, then the $\chi^2$ function will be essentially independent of
  $\Lambda$; 
\item this method provides a quantitative assessment of the potential
  size of the higher-order terms. Furthermore, an uncertainty
  arising from the truncation of the expansion is automatically
  incorporated into the uncertainties of the fit;
\item if lattice results are included from outside the PCR, then
  through the $\chi^2$, $\Lambda$ is optimised so as to give a
  best-estimate of a resummation of a subset of higher-order terms
  from beyond the working order of the chiral expansion.
\end{itemize}

\begin{figure}
\includegraphics[width=\columnwidth,angle=0]{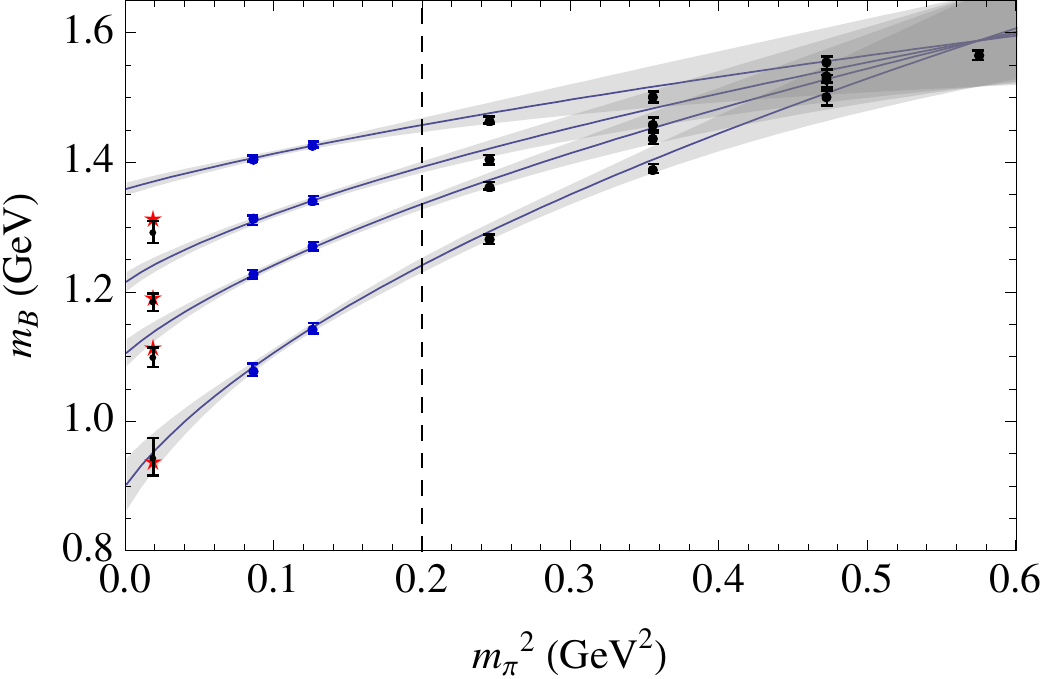}
\includegraphics[width=\columnwidth,angle=0]{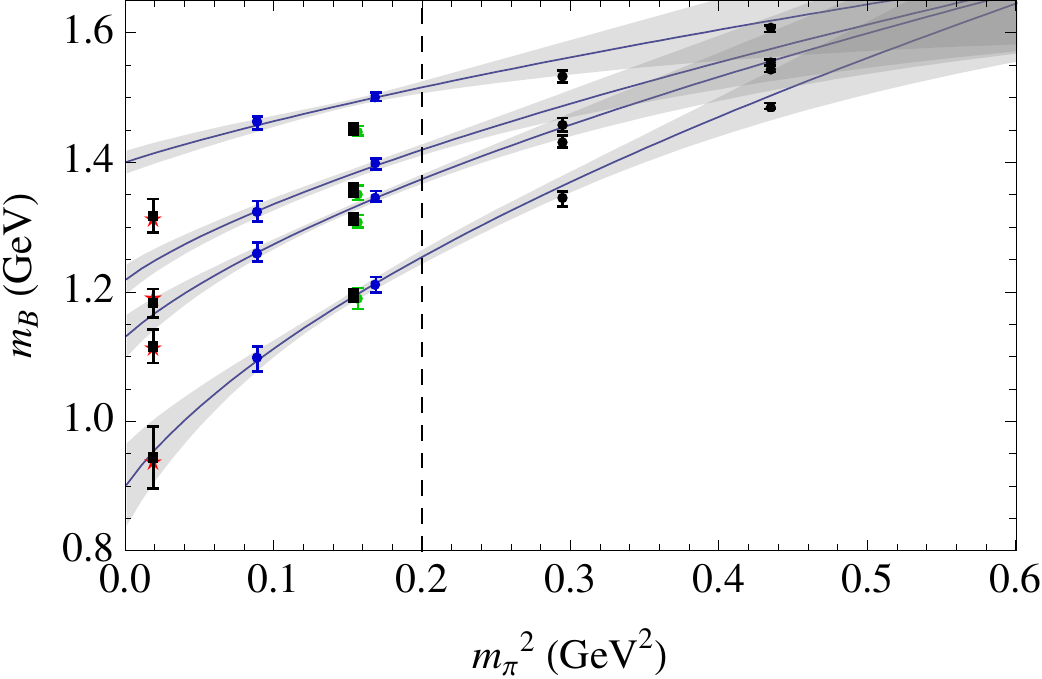}
\caption{The upper and lower panels show, respectively, the dipole
  fits (lines) to the LHPC and PACS-CS lattice simulation results
  (circles) of the octet baryon masses (curves top to bottom show
  $\Xi$, $\Sigma$, $\Lambda$ and $N$).  The squares at the physical
  pion mass display the extrapolation to the physical quark
  masses. The stars denote the physical baryon masses.  The errors
  indicated by the bands represent the total statistical errors,
  including the variation of $\Lambda$.  The third lightest pion mass
  of the PACS-CS results, which involved a lower strange quark mass,
  is {\em not} included in the displayed fit. This allows a prediction
  (squares) based on the other points (see text).}
\label{fig:LHPC}
\end{figure}

To begin, we introduce a single new parameter through the
regularisation scale. The best-fit to the LHPC (PACS-CS) results, with
5 fit parameters, is shown in the upper (lower) panel of
Fig.~\ref{fig:LHPC}, where the dipole form is shown as an example.  As
explained above, the fits include only those simulation points for
$\mpi^2<0.2\gev^2$ (the largest kaon mass is $m_K^2\simeq0.40\gev^2$).
Nevertheless, we see that the level of agreement with the lattice
simulations at higher $m_\pi^2$ is remarkably good.  Further, the
low-energy constants obtained in the two data sets are in
agreement to better than half the statistical precision of their
determination, where for example, the renormalised value, $M^{(0)}$, in the SU(3) chiral limit
is found to be $0.82\pm0.06$ and $0.83\pm0.08\gev$ for the
LHPC and PACS-CS results respectively.

The fits are determined by evaluating the baryon mass function at each
of the lattice kaon and pion masses.  The curves in
Figure~\ref{fig:LHPC} are shown for illustrative purposes, where the
kaon mass at any point on the curve is determined by fitting $m_K^2$
as a linear function in $\mpi^2$ for the corresponding lattice
ensemble. The physical masses are determined by evaluating the fit
function at the physical pion and kaon masses (which has no bearing on
the linear form used for the figure). To illustrate how this
extrapolation in the strange-quark mass works, the lower panel of
Fig.~\ref{fig:LHPC} shows a fit to just the two PACS-CS results at
fixed $\kappa_s$. Using that fit, the results of the simulation at the
different $\kappa_s$ are shown as a {\em prediction} --- by evaluating
the fit function at the lattice kaon mass of this ensemble.  The
agreement between the results of the simulation and the predicted
values is illustrative of the reliability of the fit in estimating the
dependence of the octet masses on the strange-quark mass.  In our
final results, the lattice points at this extra strange-quark mass
{\em are} also included.

For the LHPC (PACS-CS) results the optimal dipole regularisation scale
is found to be $\Lambda=1.1\pm 0.4\gev$ ($0.91\pm0.34\gev$). The
minimum $\chisq$ is 0.25 (0.05), where the improvement over the above
linear forms is evident. As discussed, the existence of a preferred
regularisation scheme is a direct signature that the results lie
outside the PCR.  Nevertheless, the extrapolated precision obtained is
rather encouraging, considering that the uncertainties incorporate the
effect of the relatively large range permitted for the regularisation
scale, as much as $\sim 0.7$--$1.5\gev$.

Given that our results do demonstrate that we are outside the PCR, we
also investigate the model-dependence in the choice of regulator. Smooth
forms, such as a dipole, monopole and a Gaussian give
indistinguishable results, with the most significant difference to
these seen in a sharp cutoff, as was seen in
Ref.~\cite{Leinweber:2003dg}.  Our estimate of the model-dependent
uncertainties incorporates the full variation over these different
functional forms, in addition to all the uncertainties in the
phenomenological input parameters, as described above.

In Table~\ref{tab:phys} we report best-estimates of the physical
masses by combining the different lattice simulations assuming they
are each good approximations of the continuum limit. We treat the
difference between the simulations as an estimate of the
discretisation uncertainty. The size of these estimates
are consistent with the findings of
Ref.~\cite{Durr:2008zz}.  We also report the dimensionless baryon
sigma terms, $\bar{\sigma}_{Bq}\equiv(m_q/M_B)\del M_B/\del m_q$. We note
that the model-dependence, is relatively small compared with the
present statistical precision. This model-dependence will not
decrease with increased statistics and therefore the present analysis,
at this order and at these quark masses, is precision
limited. Improvements beyond this will certainly require increased
efforts in the EFT and numerical computations deeper in the chiral
regime.

Our results show agreement between the absolute values of the baryon
masses and the corresponding experimental values.  This is an
important independent confirmation of the absolute scale determination reported
by Aubin~{\it et al.} \cite{Aubin:2004wf}. 
%Indeed the agreement
%appears better than one could expect based on the nominal 2\% error
%quoted in Ref.~\cite{Aubin:2004wf}, suggesting that the quoted
%precision may have been underestimated.  
Further, the sigma terms extracted by differentiating the fitting
formulae, have some quite interesting features. The pion-nucleon sigma
commutator is, within uncertainties, compatible with phenomenological
estimates \cite{Pavan:2001wz}.  The strangeness sigma commutator is
consistent with best EFT estimates~\cite{Borasoy:1996bx}, yet an order
of magnitude more precise.  This small value is observed to be
consistent with ``unquenching'' estimates~\cite{Flambaum:2004tm}, as
well as a recent 2-flavour lattice QCD estimate~\cite{Ohki:2008ff}.

In line with quark-model expectations, we note that the $\Xi$ is most
sensitive to the strange quark mass.  The linear projections of kaon
masses of the LHPC and PACS-CS results, used in Fig.~\ref{fig:LHPC},
give strange quark masses (or $m_K^2-\mpi^2/2$) that are $\sim30\%$
too high. With the definition of the sigma term giving $\frac{\delta
  M_B}{M_B}=\bar{\sigma}_{Bq}\frac{\delta m_q}{m_q}$, and 
$\bar{\sigma}_{\Xi s}\sim 0.24$ from Table~\ref{tab:phys}, this
indicates that the $\Xi$ mass on each curve should lie high by roughly
$0.24\times0.30\sim7\%$.
\begin{table}
  \caption{Extracted masses and sigma terms for the physical baryons. 
The first error is statistical; 
the second estimates the discretisation artifacts by the difference between the  
results for the LHPC and PACS-CS results; the third error represents a 
model-dependence estimate as described in the text. A further
$\sim 2\%$ uncertainty in the absolute scale~\cite{Aubin:2004wf} is understood.
The experimental masses are shown for comparison. 
\label{tab:phys}}
\begin{ruledtabular}
\begin{tabular}{l|ll|ll}
$B$       & Mass (GeV)         & Expt.   & $\bar{\sigma}_{Bl}$  & $\bar{\sigma}_{Bs}$ \\
\hline
$N$       & $0.945(24)(4)(3)$  & $0.939$ & $0.050(9)(1)(3)$    & $0.033(16)(4)(2)$  \\
$\Lambda$ & $1.103(13)(9)(3)$  & $1.116$ & $0.028(4)(1)(2)$    & $0.144(15)(10)(2)$ \\
$\Sigma$  & $1.182(11)(2)(6)$  & $1.193$ & $0.0212(27)(1)(17)$ & $0.187(15)(3)(4)$  \\
$\Xi$     & $1.301(12)(9)(1)$  & $1.318$ & $0.0100(10)(0)(4)$  & $0.244(15)(12)(2)$
\end{tabular}
\end{ruledtabular}
\end{table}
%

%Before closing, we note that it may prove possible to improve the 
%results presented here by reexpressing the expansion in 
%terms of lattice-measured quantities. This technique has been 
%demonstrated to tame the chiral expansion in other systems, 
%primarily in the mesonic sector~\cite{Beane:2005rj,Beane:2006kx}. 
%Within the framework of this manuscript,
%the clear signature of an improvement offered by using
%lattice-measured quantities would be a reduced sensitivity
%to the regularisation scale. This would be a very interesting topic for
%future work.

In summary, we have demonstrated an excellent description of current
2+1-flavour lattice QCD results based on a low-order SU(3) chiral
expansion. While the expansion to this order is not sufficiently
convergent to ensure that the fits are model-independent, we find that
the model-dependence is actually small compared with the current
statistical precision. In the future, as more simulations and
statistics accumulate, the significance of the model-dependent
component of the error will increase and will necessitate
an increased effort in the EFT.  We also anticipate that future
studies will strengthen the numerical constraints on both the continuum
and infinite-volume extrapolations, such as the work of
Ref.~\cite{Durr:2008zz}.

Within the caveats of the present study, we have demonstrated a robust
and precise determination of the absolute values of the octet bayon
masses. The controlled extrapolations have also permitted a reliable
determination of the baryon sigma terms, where we have been able to
extract an accurate determination of the $\sigma$ commutator as well
as showing that the strange-quark sigma term of the nucleon
is considerably smaller than phenomenological
estimates. As just one example of the importance of these results, we
note the significance for dark matter searches explained in
Ref.~\cite{Ellis:2008hf}.

%%%%%%%%%%%%%%%%%%%%%%%%%%%%%%%%%%%%%%%%%%%%%%%%%%%%%%%%%%%%%%%%%%%

We wish to thank J.~Arrington, S.~Beane, C.~Roberts and J.~Zanotti for
useful discussions.
This work was supported by DOE contracts 
DE-AC02-06CH11357, under which UChicago Argonne, LLC, operates Argonne
National Laboratory,
and
DE-AC05-06OR23177, under
which Jefferson Science Associates, LLC, operates Jefferson Lab.

%%%%%%%%%%%%%%%%%%%%%%%%%%%%%%%%%%%%%%%%%%%%%%%%%%%%%%%%%%%%%%%%%%%

%
\end{document}